# Perpendicular magnetic anisotropy and magnetization process in CoFeB/Pd multilayer films

Duc-The Ngo, [1,a] Duy-Truong Quach,[2] Quang-Hung Tran,[1] Kristian Mølhave,[1] The-Long Phan,[2] and Dong-Hyun Kim,[2,a]

[1]*Department of Micro-and Nanotechnology, Technical University of Denmark, Kgs. Lyngby 2800, Denmark*

[2]*Department of Physics, Chungbuk National University, Cheongju 361-763, Republic of Korea*

**Abstract**

Perpendicular magnetic anisotropy (PMA) and dynamic magnetization reversal process in [CoFeB *t* nm/Pd 1.0 nm]$_n$ (*t* = 0.4, 0.6, 0.8, 1.0, and 1.2 nm; n = 2 – 20) multilayer films have been studied by means of magnetic hysteresis and Kerr effect measurements. Strong and controllable PMA with an effective uniaxial anisotropy up to $7.7 \times 10^6$ J.m$^{-3}$ and a saturation magnetization as low as 200 emu/cc are achieved. Surface/interfacial anisotropy of CoFeB/Pd interfaces, the main contribution to the PMA, is separated from the effective uniaxial anisotropy of the films, and appears to increase with the number of the CoFeB/Pd bilayers. Observation of the magnetic domains during a magnetization reversal process using polar magneto-optical Kerr microscopy shows the detailed behavior of nucleation and displacement of the domain walls.

[a] Authors to whom correspondence should be addressed. Electronic addresses: dngo@nanotech.dtu.dk and donghyun@chungbuk.ac.kr

1. INTRODUCTION

Perpendicular magnetic anisotropy (PMA) thin films play a crucial role in emerging spin-electronic technology using the spin-transfer torque phenomenon.[1-3] With magnetic moments aligned perpendicular to the film plane, the PMA is committed to enhance the spin-switching efficiency that allows significant reduction of the current density magnetization in such systems. Many thin films with PMA have been investigated for spin-transfer torque applications, such as, Co-based multilayers *e.g.* Co/Pt,[2,4], Co/Pd,[5,6] CoFe/Pd,[7,8] $L_{10}$ single layer films, *e.g.* MnGa,[9,10] CoPt, FePt films,[11,12] CoFeB/MgO films,[3,13-15] and rare earth – transition metal amorphous films *e.g.* TbFeCo,[16-18] GdFeCo,[19] *etc.* Among them, CoFeB-based films are attractive because of high spin polarization,[20] and moderate saturation magnetization,[13-15] which are essential for reducing the current density of spin-switching. Commonly, efficient formation of PMA in crystalline CoFeB films requires *bcc*-CoFeB with (001) texture (perpendicular to film plane) and MgO/CoFeB interface.[13,15] Thus, in-situ heating during deposition or post annealing is commonly a demand to convert amorphous film into the required crystal structure.[21,22] However, heat treatment of the films with oxide interfaces would easily result in the oxidation and migration (of Boron, Oxygen) that influence on the magnetic properties of the films.[23-25]

Recently, PMA in multilayer films using amorphous CoFeB layers and noble metal spacing have been reported.[26-28] Several advantages allow considering PMA amorphous CoFeB films as an excellent candidate for spin-transfer torque applications:

- Importantly, the CoFeB amorphous films possess high spin polarization (over 65%),[20] even higher than that in crystalline CoFeB film.[29]

- The absence of grain boundaries in an amorphous thin layer should significantly reduce the pinning site density that hinders the magnetization reversal. This can enhance the speed of spin-switching in spin-transfer torque devices. Beside that amorphous layers could create very smooth interfaces those reduce the attenuation of the electron spin as transfers through the interface, desirable for high tunneling magnetoresistance,[30,31] if being used in magnetic tunneling junctions.

- Multilayer structure of amorphous CoFeB and nobble metal spacing (*e.g.* Pd, Pt) avoids the risk of oxidation and diffusion those would affect to the magnetic properties of the films.

- The presence of B atoms with smaller radius than those of Co and Fe results in a high glass forming ability,[32] and consequently, high stability as compared to rare-earth transition metal amorphous films (*e.g.* TbFeCo, GdFeCo)[16-19] is expected even under heating.

- Moreover, the PMA in amorphous CoFeB films could occur easily without any further heat treatment and with any kind of common substrates. From the technical opinion, this makes the fabrication process simple and less expensive.

This paper presents a systematic study of perpendicular magnetic anisotropy and the detailed magnetization process in terms of in-situ observation of magnetic domain structure in the series of amorphous CoFeB multilayer films with changing of magnetic layer thickness as well as number of CoFeB/Pd bilayers, in which the noble metal Pd is used as spacing layers. The contribution of interfacial magnetization and anisotropy to the PMA of the films is separated from the variation of effective uniaxial anisotropy with the CoFeB thickness, which plays an essential role in the formation of PMA in such CoFeB-based multilayers.

## 2. EXPERIMENTS

The multilayer films with stacks of Sub/Ta 2.0/Pd 2.0/(CoFeB $t$/Pd 1.0)$_n$/Ta 1.0 (thickness in nm; $t$ = 0.4 – 1.2 nm, and n = 2, 4, .. 20) were grown on Si wafers with a native oxide layer by using a DC magnetron sputtering system (ULVAC UHV). The base pressure was better than $5.0 \times 10^{-9}$ Torr and the plasma Ar pressure was remained at 1.0 mTorr. The sputter power was fixed at 30 W for the CoFeB target, 25 W for Pd target, and 50 W for Ta to ensure very low deposition rate (~0.01-0.02 nm/s). Nominal composition of the CoFeB target was 40:40:20 as this is expected to possess the highest spin polarization among the amorphous ferromagnetic Co-Fe-B alloys.[20] The underlayer Ta 2.0/Pd 2.0 nm was selected to work as buffer layer for good adhesion and good surface morphology. Atomic force microscopy (AFM) imaging (data not shown) revealed a root-mean-square surface roughness of the Ta/Pd buffer layer of 0.20 nm. The top Ta layer (1.0 nm thick) was used as a capped layer to protect the films from ambient oxidation. Magnetic properties of the films were measured using a vibrating sample magnetometer (VSM) with a maximum field of 2.0 T (Lakeshore 7400 VSM). Magnetization reversal processes in the films were observed in terms of magnetic domain by polar Kerr microscopy with an out-of-plane field up to 140 mT.[33]

## 3. RESULTS AND DISCUSSION

Fig. 1 shows the magnetic hysteresis (*M-H*) loops of CoFeB/Pd multilayers with variation of CoFeB thickness (a) and variation of number of CoFeB/Pd bilayers (b). The inset of Fig. 1(a) shows the *M-H* loops as a typical sample (CoFeB 0.4nm/Pd 1.0nm)$_{10}$ measured in-plane and out-of-plane of the film. Fig. 1(a) indicates that the PMA has established by a rectangular hysteresis loop in the thin films with low thickness of CoFeB layer (lower than 1.2 nm) as measured with the applied field perpendicular to the film plane. The magnetization appears to align in-plane of the film when the thickness reaches 1.2 nm. It is interesting to note from Fig. 1(b) that the PMA still exist in the multilayer with 20 bilayers of CoFeB/Pd with the CoFeB layers of 0.4 nm.

From the hysteresis loops, two most important magnetic parameters including saturation magnetization, $M_s$, and effective uniaxial anisotropy, $K_{eff}$, are deduced. By overlapping the hysteresis loops measured on two directions (in-plane and out-of-plane), the saturation field, $H_s$, is determined as the intercept of two loops, see insert in fig 1(a). As a result, the anisotropy field, $H_k$, is obtained as:[8,14]

$$H_k = H_s + 4\pi M_s \tag{1}$$

and the effective uniaxial anisotropy can be calculated by:[8,14]

$$K_{eff} = H_k \times M_s/2 \qquad (2)$$

Fig. 2(a) illustrates the $M_s$ as a function of the CoFeB layer thickness, $t$, for the thin films with 10 CoFeB/Pd bilayers. Obviously, the saturation magnetization increases with the CoFeB thickness, from 400 emu/cc to 820 emu/cc, as a linear function similar to a previous study on a PMA CoFeB film with MgO interfaces.[14] This is ascribed to the increase of the volume magnetic moment when increasing the thickness of the CoFeB layer. Increasing the number of CoFeB/Pd bilayers [detailed data are not shown here, but an example can be seen in Fig. 1(b)] also shows a similar tendency in the saturation magnetization, that looks similar to the previous results on amorphous CoFeB multilayers published elsewhere.[21] The lowest $M_s$ value of 200±28 emu/cc is found in the sample with 2 bilayers of CoFeB/Pd at a CoFeB thickness of 0.4 nm. Please note that a strong PMA film with such a low $M_s$ is desirable for spin-transfer torque devices as this will likely reduce the spin-switching current density,[7-10,16-18] which is supposed to be dependent on the saturation magnetization of the spin-polarizing layers. Qualitatively, the magnetization in the films could be considered as two terms: volume magnetization and surface/interface magnetization. As the number of CoFeB/Pd bilayers is fixed, and the CoFeB thickness is changed, the volume term would vary whereas the surface/interfacial term would assume to be unchanged. Thus, the surface/interfacial magnetization could be estimated from the linear dependence of the total magnetization of the CoFeB. By extrapolating the linear plot to zero thickness (the volume term reduced to zero), this zero-thickness magnetization will represent the surface/interfacial term. In the case given in Fig. 2(a), a value of 180 emu/cc is deduced and attributed to the magnetization of the CoFeB/Pd interfaces. This value is nearly invariable when the number of CoFeB/Pd bilayers increases from 2 to 20 bilayers.

As fixing the number of CoFeB/Pd bilayers, the effective uniaxial anisotropy, $K_{eff}$, decreases with increasing CoFeB thickness. Namely, for the multilayers with n = 10, the $K_{eff}$ decreases from $7.7\times10^6$ J.m$^{-3}$ for $t$ = 0.4 nm, to $1.0\times10^6$ J.m$^{-3}$ for $t$ = 1.0 nm, and to $-3.2\times10^5$ J.m$^{-3}$ for $t$ = 1.2 nm. It is noticed that the negative sign of effective anisotropy denotes the in-plane anisotropy existing in the film. This variation can be understood by considering the effective anisotropy as two terms: volume anisotropy, $K_v$ and surface/interface anisotropy, $K_s$.[34] The $K_v$, contributed by the volume magnetic moment, is negative and tends to align the magnetization in-plane of the film. In contrast, the $K_s$, caused by the surface and interfacial magnetic moments, is positive and directs the magnetization perpendicular to the film plane. Competition between these terms determines the sign of the effective anisotropy: positive for out-of-plane anisotropy, and negative for in-plane anisotropy. The relation between these terms is expressed by the equation:[34]

$$K_{eff} = K_v + 2K_s/t \qquad (3)$$

where $t$ is the thickness of the CoFeB layer. This relation could be rewritten as:

$$K_{eff} \cdot t = K_v \cdot t + 2K_s \qquad (4)$$

From (4), the anisotropy terms, $K_s$ and $K_v$ can be separated from the $K_{eff}$ in terms of the linear dependence of $K_{eff} \cdot t$ on the thickness ($t$). Fig. 2(b) plots the linear dependence of $K_{eff} \cdot t$ on the thickness ($t$) for the multilayer system with n = 10, and CoFeB thickness varying from 0.4 - 1.2 nm. The $2K_s$ is determined as the intercept of the linear dependence with the horizontal axis whereas the $K_v$ is the slope of the linear dependence. In this case, a $K_s$ value of $2.44\times10^{-3}$ Jm$^{-2}$ and a $K_v$ value of $-4.16\times10^{-3}$ Jm$^{-3}$ are deduced from the plot. This also allows determining the critical thickness of CoFeB layer, for which the magnetic anisotropy switches from perpendicular to in-plane, as $t_c = -K_s/K_v = 1.17\pm0.01$ nm. Applying this approach to all the other multilayers with different numbers of CoFeB/Pd bilayers, the surface/interfacial anisotropy is found to be varies as a function of the CoFeB/Pd bilayer numbers, which is shown in Fig. 3 as a linear function. The increase in the number of CoFeB/Pd bilayers results in the increase of the CoFeB/Pd interfaces, and this gives rise to an enhancement of the interfacial anisotropy as observed. All these quantifications of the surface/interfacial anisotropy and interfacial component of the $M_s$ confirms that the PMA in the studied CoFeB/Pd multilayers predominantly originates from the surface/interfacial magnetic moments of the CoFeB/Pd interfaces due to the broken symmetry of ultrathin layer structure.[35] Additionally, previous studies suggested that these magnetic moments might arise from the magnetic polarization of Pd 4d caused by hybridization with the Co and Fe 3d ions.[8,36] Additionally, the use of nobble metal Pd as spacing layers could bring more benefit to the film if being used in spin-electronic devices because of the long spin-diffusion length of Pd,[37,38] and less structural diffusion to magnetic layers to make a stable PMA owning to high melting temperature of Pd.

To understand the magnetization reversal mechanism of the films, in-situ observation of the magnetic domain structure in the films during a magnetization reversal process was performed by means of polar Kerr effect microscopy supplemented with hysteresis loops measurements.[33] Fig. 4 presents a reversal process in the multilayer sample [CoFeB/Pd]$_6$ with a CoFeB layer thickness of 0.4 nm. The reversal starts as reversing the out-of-plane field from the maximum positive field to a negative field of -15.0 mT. At this point, reversed domains with opposite direction of magnetization denoted by dark spots on the bright background [Fig. 4(a)] are nucleated. When slightly increases the negative field to -17.0 mT, these domains start growing in the opposite direction as seen in [Fig. 4(b)], where the dark spots increase their size quickly. Continue increasing the negative field to the coercive field (-17.5 mT) makes the dark-contrast reversed domains increasing their size (domain walls are propagating). At this field, net zero-magnetization could be visible via the same areas of the bright domains (positive direction of the magnetization) and the dark domains (negative direction of the magnetization). When the negative field increases slightly further (~18.0 mT), the dark region expands to be major in the film, and the reversal process mostly completes [Fig. 4(d,e)]. A square-shaped hysteresis loop [Fig. 4(f)] is also deduced from the MOKE images and shown to be similar to that obtained by the VSM.

The variation of domain structure here shows a typical reversal process on the easy axis, in which the reversal is governed by the nucleation of magnetic domains and domain wall movement. It is noteworthy that the growth of the domains and the movement of the domain walls during the reversal seem to be steady, indicating a weak pining effect on the propagation of the walls as expected in the amorphous film. The domain structure in the investigated amorphous CoFeB-based multilayer films looks totally different from other cubic-CoFeB films with PMA in which stripe domain structure was observed.[39] This aspect would be very promising for application in domain-wall-controlled spin-torque devices. Similar domain structure and reversal behavior are also observed in other samples with PMA. They exhibit similar to the Co/Pt multilayers observed previously.[40]

Furthermore, the nucleation of the reversed domains during the reversal seems to be changed by the number of the CoFeB/Pd bilayers. Fig. 5 depicts the domain patterns of some representative samples (the thickness of the CoFeB layers here is 0.4 nm, as a typical example) at the states where the reversed domains are nucleated. Apparently, the number of reversed domains increases with the number of the CoFeB/Pd bilayers, which determines the relative importance of interfaces in the films. This suggests that the reversed domains are essentially nucleated at the interfaces of which interfacial roughness would play an important role of such a nucleation.

## 4. CONCLUSIONS and OUTLOOKS

We presented a systematic study of perpendicular magnetic anisotropy in amorphous CoFeB/Pd multilayer films. Strong PMA can be established in a large range of CoFeB thickness up to 1.0 nm with controllably low saturation magnetization (down to 200 emu/cc), and high effective uniaxial anisotropy (up to ~$7.7 \times 10^6$ J.m$^{-3}$). The PMA vanishes when increasing the CoFeB layer thickness to 1.2 nm. By varying the CoFeB layer thickness and the number of (CoFeB/Pd) bilayers, the contribution of volume and surface/interface magnetization and anisotropy to the magnetism of the films are separated and quantified. A magnetization of the interfacial magnetic moment is estimated to be 180±28 emu/cc. The strong PMA of the films is essentially contributed by the surface/interface anisotropy. In-situ observation of magnetization reversal process in the films using polar magneto-optical Kerr microscopy indicates the nucleation of domains and domain wall movement mechanism in the studied films in which the CoFeB/Pd interfaces play an important role for nucleation of reversed domains.

The studied multilayer films using amorphous CoFeB layers are excellent candidates for applications ranging from spin-transfer torque to conventional perpendicular magnetic recording. With high out-of-plane anisotropy, very thin Bloch-type walls are expected to exist in the films (the wall width can be estimated to be sub-5 nm using the relation $\Delta \approx \pi \sqrt{A/K_u}$,[8] with $A \approx 20$ pJ/m for CoFeB-based films).[39] By adjusting the thickness of the layer thickness and the number of CoFeB/Pd bilayers, the $M_s$ can be reproducibly controlled from very low range (~280 emu/cc) to moderately high 830 emu/cc. Additionally, the amorphous CoFeB films would be expected to have a very low Gilbert damping coefficient down to 0.012 due to the homogeneity of the interfaces.[41] Low $M_s$, high $K_u$ (very thin Bloch wall width) and low damping coefficient are desirable for creating spin-transfer torque devices working with low current density of spin-switching to compete with the other PMA thin films.[13,42] Regarding the magnetization reversal, the studied multilayer films would be very potential to apply to the domain-wall spin-torque devices, particularly, because the interfacial-supported domain nucleation and steady movement of the domain walls would be fruitful for simple control of domain walls motion. Moderately high $M_s$ in some samples would be an interesting aspect for perpendicular magnetic recording technology.[43]

Generally, the requirement of *fcc* structure with (111) texture perpendicular to the film plane in the layers of Co-based (or CoFe-based) multilayers has been proved to be the first condition for the formation of PMA.[2-8] However, the results in this article (and some recent papers)[26-28] have demonstrated that the structural condition is not a necessary criterion for the CoFeB/Pd interfaces. This is presumably a physically interesting aspect for a deep analysis of interfacial magnetic moment and electronic structure of the CoFeB/Pd interfaces by a number of tools, *e.g.* X-ray magnetic circular dischroism,[44,45] neutron diffraction.[46,47]

## ACKNOWLEDGEMENTS

This work was supported by the Leading Foreign Research Institute Recruitment Program under Grant 2012K1A4A3053565 and Grant 2010-0021735 through the National Research Foundation of Korea.

**REFERENCES**

[1] S. Mangin, D. Ravelosona, J. A. Katine, M. J. Carey, B. D. Terris, and E. E. Fullerton, Nat. Mater. **5**, 210 (2006).

[2] O. Boulle, J. Kimling, P. Warnicke, M. Kläui, U. Rüdiger, G. Malinowski, H. J. M. Swagten, B. Koopmans, C. Ulysse, and G. Faini, Phys. Rev. Lett. **101**, 216601 (2008).

[3] S. Yakata, H. Kubota, Y. Suzuki, K. Yakushiji, A. Fukushima, S. Yuasa, and K. Ando, J. Appl. Phys. **105**, 07D131 (2009).

[4] S. Bandiera, R. C. Sousa, B. Rodmacq, and B. Dieny, Appl. Phys. Lett. **100**, 142410 (2012).

[5] M. Jamali, K. Narayanapillai, X. Qiu, L. M. Loong, A. Manchon, and H. Yang, Phys. Rev. Lett. **111**, 246602 (2013).

[6] S. Hashimoto, Y. Ochiai, and K. Aso, J. Appl. Phys. **66**, 4909 (1989).

[7] J. Bae, H.-J. Kim, J. Chang, S. H. Han, H. C. Koo, and S. H. Lim, J. Kor. Phys. Soc. **61**, 1500 (2012).

[8] D.-T. Ngo, Z. L. Meng, T. Tahmasebi, X. Yu, E. Thoeng, L.H. Yeo, A. Rusydi, G. C Han, K.-L. Teo, J. Magn. Magn. Mater. **350**, 42 (2014).

[9] H. Kurt, K. Rode, M. Venkatesan, P. S. Stamenov, and J. M. D. Coey, Phys. Rev. B **83**, 020405 (2011).

[10] S. Mizukami, T. Kubota, F. Wu, X. Zhang, T. Miyazaki, H. Naganuma, M. Oogane, A. Sakuma, and Y. Ando, Phys. Rev. B **85**, 014416 (2012).

[11] J.-U. Thiele, L. Folks, M. F. Toney, and D. K. Weller, J. Appl. Phys. **84**, 5686 (1998).

[12] C.-J. Lin, and G. L. Gorman, Appl. Phys. Lett. **61**, 1600 (1992).

[13] S. Ikeda, K. Miura, H. Yamamoto, K. Mizunuma, H. D. Gan, M. Endo, S. Kanai, J. Hayakawa, F. Matsukura, and H. Ohno, Nat. Mater. **9**, 721 (2010).

[14] M. Endo, S. Kanai, S. Ikeda, F. Matsukura, and H. Ohno, Appl. Phys. Lett. **96**, 212503 (2010)

[15] S. Fukami, T. Suzuki, Y. Nakatani, N. Ishiwata, M. Yamanouchi, S. Ikeda, N. Kasai, and H. Ohno, Appl. Phys. Lett. **98**, 082504 (2011).

[16] S. Li, H. Nakamura, T. Kanazawa, X. Liu, and A. Morisako, IEEE Trans. Magn. **46**, 1695 (2010).

[17] D.-T. Ngo, K. Ikeda, and H. Awano, Appl. Phys. Express **4**, 093002 (2011).

[18] D.-T. Ngo, K. Ikeda, and H. Awano, J. Appl. Phys. **111**, 083921 (2012).

[19] M. Ding, S. J. Poon, J. Magn. Magn. Mater. **339**, 51 (2013).

[20] S. X. Huang, T. Y. Chen, and C. L. Chien, Appl. Phys. Lett. **92**, 242509 (2008).

[21] K. Mizunuma, S. Ikeda, J. H. Park, H. Yamamoto, H. Gan, K. Miura, H. Hasegawa, J. Hayakawa, F. Matsukura, and H. Ohno, Appl. Phys. Lett. **95**, 232516 (2009).

[22] H. Meng, W. H. Lum, R. Sbiaa, S. Y. H. Lua, and H. K. Tan, J. Appl. Phys. **110**, 033904 (2011).

[23] J. Y. Bae, W. C. Lim, H. J. Kim, T. D. Lee, K. W. Kim, and T. W. Kim, J. Appl. Phys. **99**, 08T316 (2006).

[24] A. K. Rumaiz, C. Jaye, J. C. Woicik, W. Wang, D. A. Fischer, J. Jordan-Sweet, and C. L. Chien, Appl. Phys. Lett. **99**, 222502 (2011).

[25] Y.-P. Wang, G.-C. Han, H. Lu, J. Qiu, Q.-J. Yap, R. Ji, and K.-L. Teo, J. Appl. Phys. **114**, 013910 (2013).

[26] J. H. Jung, B. Jeong, S. H. Lim, and S.-R. Lee, Appl. Phys. Express **3**, 023001 (2010).

[27] C. Fowley, N. Decorde, K. Oguz, K. Rode, H. Kurt, and J. M. D. Coey, IEEE Trans. Magn. **46**, 2116 (2010).

[28] J. H. Jung, S. H. Lim, and S.-R. Lee, J. Nanosci. Nanotech. **11**, 6233 (2011).

[29] P. V. Paluskar, J. J. Attema, G. A. de Wijs, S. Fiddy, E. Snoeck, J. T. Kohlhepp, H. J. M. Swagten, R. A. de Groot, and B. Koopmans, Phys. Rev. Lett. **100**, 057205 (2008).

[30] D. Wang, C. Nordman, J. M. Daughton, Z. Qian, and J. Fink, IEEE Trans. Magn. **40**, 2269 (2004).

[31] A. T. G. Pym, A. Lamperti, B. K. Tanner, T. Dimopoulos, M. Rührig, and J. Wecker, Appl. Phys. Lett. **88**, 162505 (2006).

[32] Y. Luo, M. Esseling, A. Käufler, K. Samwer, T. Dimopoulos, G. Gieres, M. Vieth, M. Rührig, J. Wecker, C. Rudolf, T. Niermann, and M. Seibt, Phys. Rev. B **72**, 014426 (2005).

[33] S.-C. Shin, K.-S. Ryu, D.-H. Kim, and H. Akinaga, J. Appl. Phys. **103**, 07D907 (2008).

[34] M. T. Johnson, P. J. H. Bloemen, F. J. A. den Broeder, and J. J. de Vries, Rep. Prog. Phys. **59**, 1409 (1996).

[35] M. L. Néel, J. de Phys. et le Rad. **15**, 225 (1954).

[36] H. Sakurai, F. Itoh, Y. Okabe, H. Oike, and H. Hashimoto, J. Magn. Magn. Mater. **198-199**, 662 (1999).

[37] J. Foros, G. Woltersdorf, B. Heinrich, and A. Brataas, J. Appl. Phys. **97**, 10A714 (2005).

[38] K. Kondou, H. Sukegawa, S. Mitani, K. Tsukagoshi, and S. Kasai, Appl. Phys. Express **5**, 073002 (2012).

[39] M. Yamanouchi, A. Jander, P. Dhagat, S. Ikeda, F. Matsukura, and H. Ohno, IEEE Magn. Lett. **2**, 3000304 (2011).

[40] S.-B. Choe, and S.-C. Shin, Phys. Rev. B **57**, 1085 (1998).

[41] T. Devolder, P.-H. Ducrot, J.-P. Adam, I. Barisic, N. Vernier, J.-V. Kim, B. Ockert, and D. Ravelosona, Appl. Phys. Lett. **102**, 022407 (2013).

[42] H. Szambolics, J.-ChToussaint, A. Marty, I. M. Miron, L. D. Buda-Prejbeanu, J. Magn. Magn. Mater. **321**, 1912 (2009).

[43] J. K. W. Yang, Y. Chen, T. Huang, H. Duan, N. Thiyagarajah, H. K. Hui, S. H. Leong, and V. Ng, Nanotechnology **22**, 385301 (2011).

[44] S. Tsunegi, Y. Sakuraba, K. Amemiya, M. Sakamaki, E. Ozawa, A. Sakuma, K. Takanashi, and Y. Ando, Phys. Rev. B **85**, 180408 (2012).

[45] H. Yang, S.-H. Yang, D.-C. Qi, A. Rusydi, H. Kawai, M. Saeys, T. Leo, D. J. Smith, and S. S. P. Parkin, Phys. Rev. Lett. **106**, 167201 (2011).

[46] Y. Endoh, J. Phys. Colloques **43**, C7-159 (1982).

[47] S. Singh, S. Basu, M. Gupta, M. Vedpathakz, and R. H. Kodama, J. Appl. Phys. **101**, 033913 (2007).

**Figures**

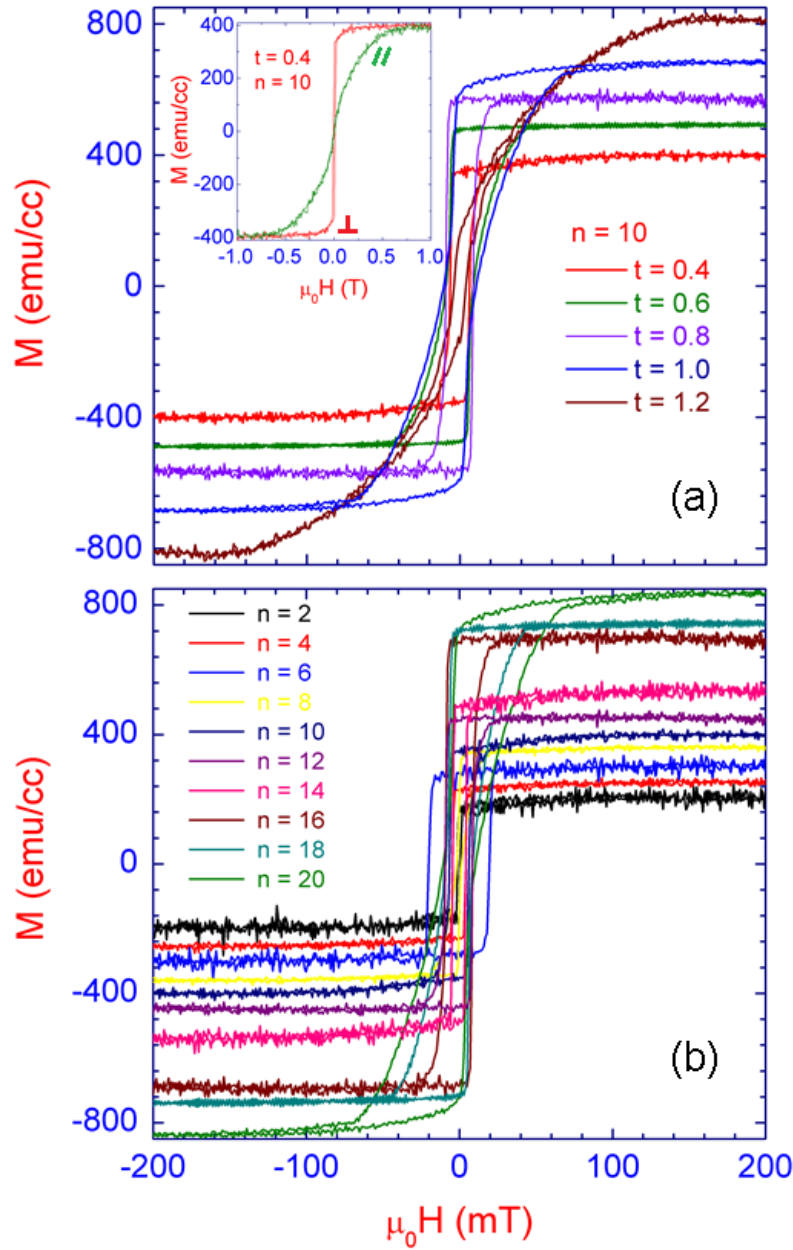

FIG. 1. (a) Hysteresis loops of the [CoFeB $t$/Pd 1.0nm]$_{10}$ films ($t$ = 0.2 – 1.2 nm) measured with magnetic field applied perpendicular to the film plane. The inset shows the hysteresis loops of the sample with $t$ = 0.4 nm with applied magnetic field in both directions: parallel and perpendicular to the film plane. (b) Hysteresis loops of the [CoFeB 0.4nm/Pd 1.0nm]$_n$ films ($n$ = 2 – 20) measured with perpendicular magnetic field.

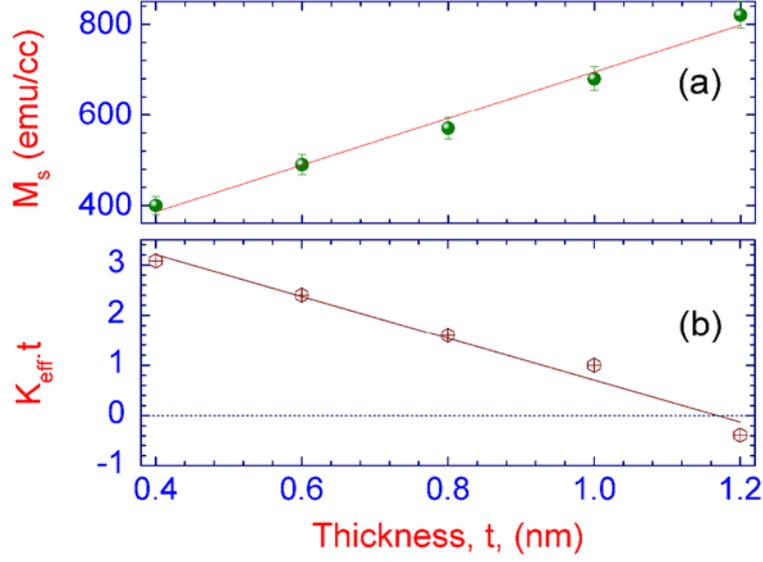

FIG. 2. (a) Saturation magnetization, $M_s$, and (b) $K_{eff} \times t$ as a function of the CoFeB layer thickness, $t$ for a multilayer system of [CoFeB $t$/Pd 1.0nm]$_{10}$ ($t = 0.2 – 1.2$ nm).

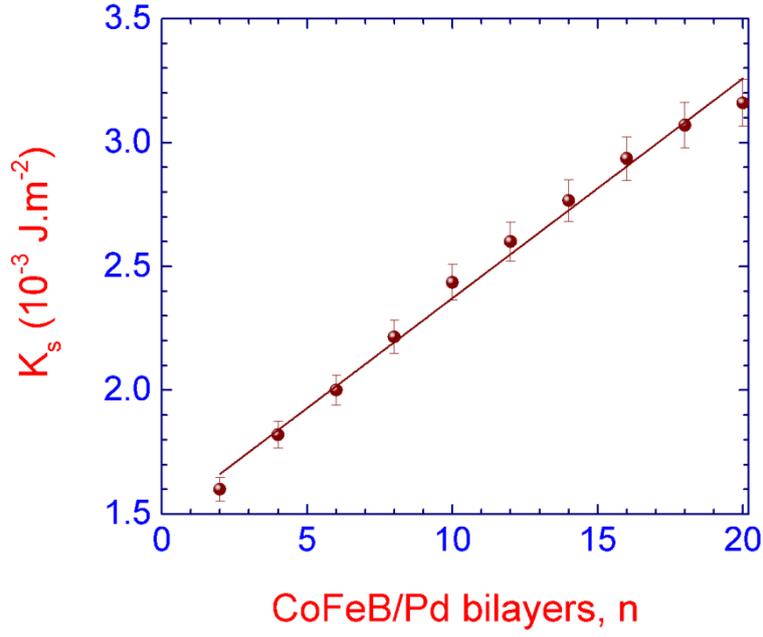

FIG. 3. Surface/interfacial anisotropy, $K_s$, extracted from the effective uniaxial anisotropy as a function of the number of CoFeB/Pd bilayers.

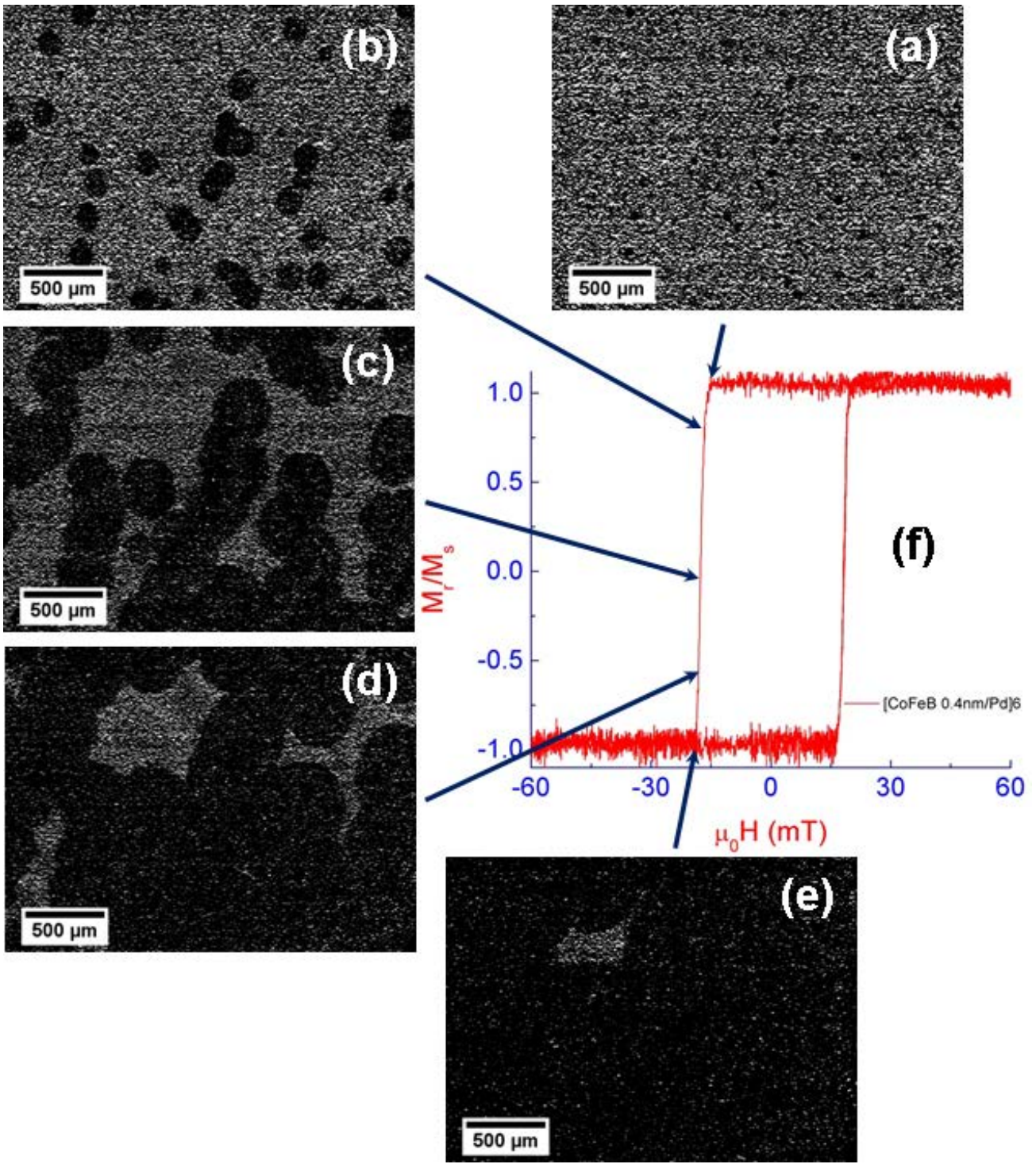

FIG. 4. Magnetization reversal process in the sample [CoFeB $t$/Pd 1.0nm]$_6$ ($t$ = 0.4 nm) observed by polar Kerr microscopy.

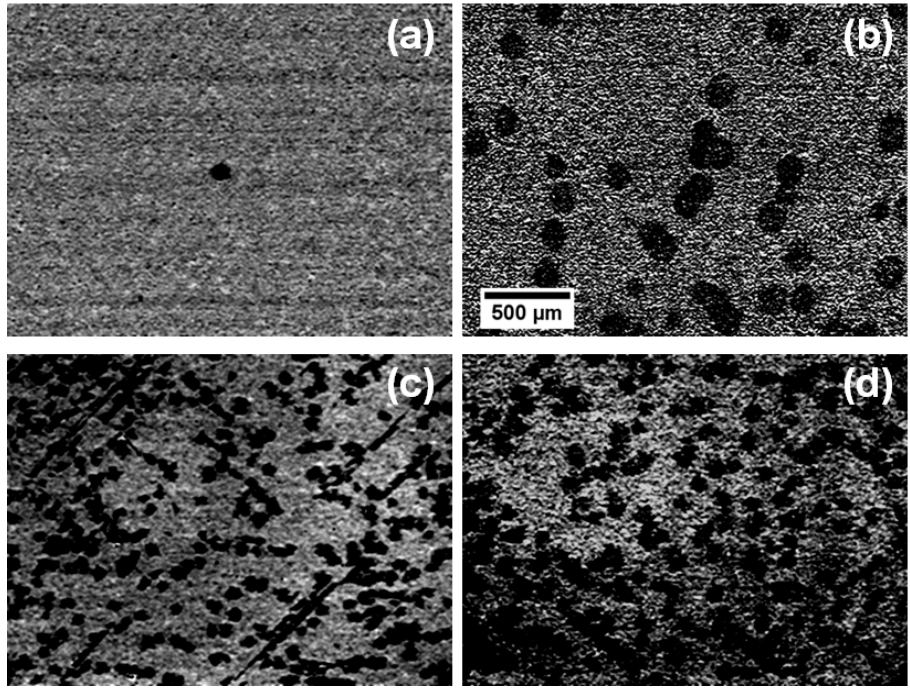

FIG. 5. Domain structure of some samples [CoFeB 0.4nm/Pd 1.0nm]$_n$ at the nucleation of reversed domains state: (a) n = 2, (b) n = 6, (c) n = 16, and (d) n = 20. All pictures have same dimension and scale bar.